\documentclass[letterpaper, 10 pt, conference]{ieeeconf}  

\usepackage{graphicx}  
\usepackage{amsmath} 
\usepackage{footnote}
\usepackage{amsmath, amssymb, amsfonts}   
\usepackage{enumerate}
\usepackage{amsmath} 
\usepackage{etoolbox} 

\usepackage{algorithm} 
\usepackage{algpseudocode} 

\let\labelindent\relax 
\usepackage{enumitem} 

\usepackage{hyperref}

\usepackage{tabularx} 
\usepackage{multirow} 
\usepackage{array}    
\usepackage{booktabs}
\newcolumntype{Y}{>{\centering\arraybackslash}X} 

\newcommand{\rtot}{R_{\mathrm{tot}}}

\newcommand{\itot}{I}

\usepackage{amsmath}
\IEEEoverridecommandlockouts                              

\title{\LARGE \bf
Differential Voltage Analysis and Patterns
in \\ Parallel-Connected Pairs of Imbalanced Cells
}

\author{Clement Wong$^{1}$, Andrew Weng$^{1}$, Sravan Pannala$^{1}$, Jeesoon Choi$^{2}$, Jason B. Siegel$^{1}$, Anna Stefanopoulou$^{1}$
\thanks{*This work was supported by LG Energy Solutions.}
\thanks{$^{1}$Clement Wong,  Andrew Weng, Sravan Pannala, Jason B. Siegel, Anna Stefanopoulou are with Department of Mechanical Engineering, University of Michigan, Ann Arbor, Michigan, 48105. 
        {\tt\small {\{clemwong,asweng,spannala,\newline siegeljb,annastef\}}@umich.edu}}%
\thanks{$^{2}$Jeesoon Choi is with LG Energy Solutions, Korea
        {\tt\small jeesoonchoi@lgensol.com}}%
}

\begin{document}

\maketitle
\thispagestyle{empty}
\pagestyle{empty}

\begin{abstract}

Diagnosing imbalances in capacity and resistance within parallel-connected cells in battery packs is critical for battery management and fault detection, but it is challenging given that individual currents flowing into each cell are often unmeasured. This work introduces a novel method useful for identifying imbalances in capacity and resistance within a pair of parallel-connected cells using only voltage and current measurements from the pair. Our method utilizes differential voltage analysis (DVA) when the pair is under constant current discharge and demonstrates that features of the pair’s differential voltage curve (dV/dQ), namely its mid-to-high SOC dV/dQ peak's height and skewness, are sensitive to imbalances in capacity and resistance. We analyze and explain how and why these dV/dQ peak shape features change in response to these imbalances, highlighting that the underlying current imbalance dynamics resulting from these imbalances contribute to these changes. Ultimately, we demonstrate that dV/dQ peak shape features can identify the product of capacity imbalance and resistance imbalance, but cannot uniquely identify the imbalances. This work lays the groundwork for identifying imbalances in capacity and resistance in parallel-connected cell groups in battery packs, where commonly only a single current sensor is placed for each parallel cell group.

\end{abstract}

\section{INTRODUCTION}

Individual batteries that make up battery modules in parallel or in series can degrade at different rates, due to factors such as manufacturing process variations that can affect the fabrication of the batteries’ electrodes \cite{Weng2023-lj, Kenney2012-hl} or module-level environmental variations in the wiring resistance, welding process, and connectors \cite{MA2018653}. Such variations in degradation among the individual batteries within a module can compound and jeopardize the entire module’s safety and performance \cite{en14113276}. Thus, for maintaining safety and performance, it is important to gauge the variability in degradation across batteries within a module. 

Existing literature outlines various methods for assessing degradation of Li-ion batteries. Two widely-used techniques are Differential Voltage Analysis (DVA) and Incremental Capacity Analysis (ICA), which leverage differential voltage (dV/dQ) and differential capacity (dQ/dV) measurements to gain insights into half-cell behavior and understand degradation mechanisms. Features from the dV/dQ and dQ/dV curves, particularly their peaks, can be harnessed as quantitative metrics for assessing the extent of degradation within a cell. By comparing the dV/dQ and dQ/dV peaks of different cells, one can identify variability in degradation among cells \cite{WENG201336,BERECIBAR2016239,Sieg2020,Lewerenz2017,FENG2020100051}. 

While methods exist for assessing degradation in a single cell and can be used to evaluate variations in degradation among cells, these methods often require individual sensors for each cell.  Manufacturers of battery modules often limit the number of sensors in a battery module to reduce cost\cite{Lin2020-tb}. In particular, for parallel-connected cells, manufacturers use only a single current sensor for each group of parallel-connected cells, leaving the individual cell currents unknown \cite{Weng2022-pt, Zhang2020-ci}. Due to this constraint, existing methods are generally not directly applicable for identifying variability in degradation among parallel-connected cells.  

\begin{figure}[ht!]
    \centering
    \includegraphics[width=\linewidth]{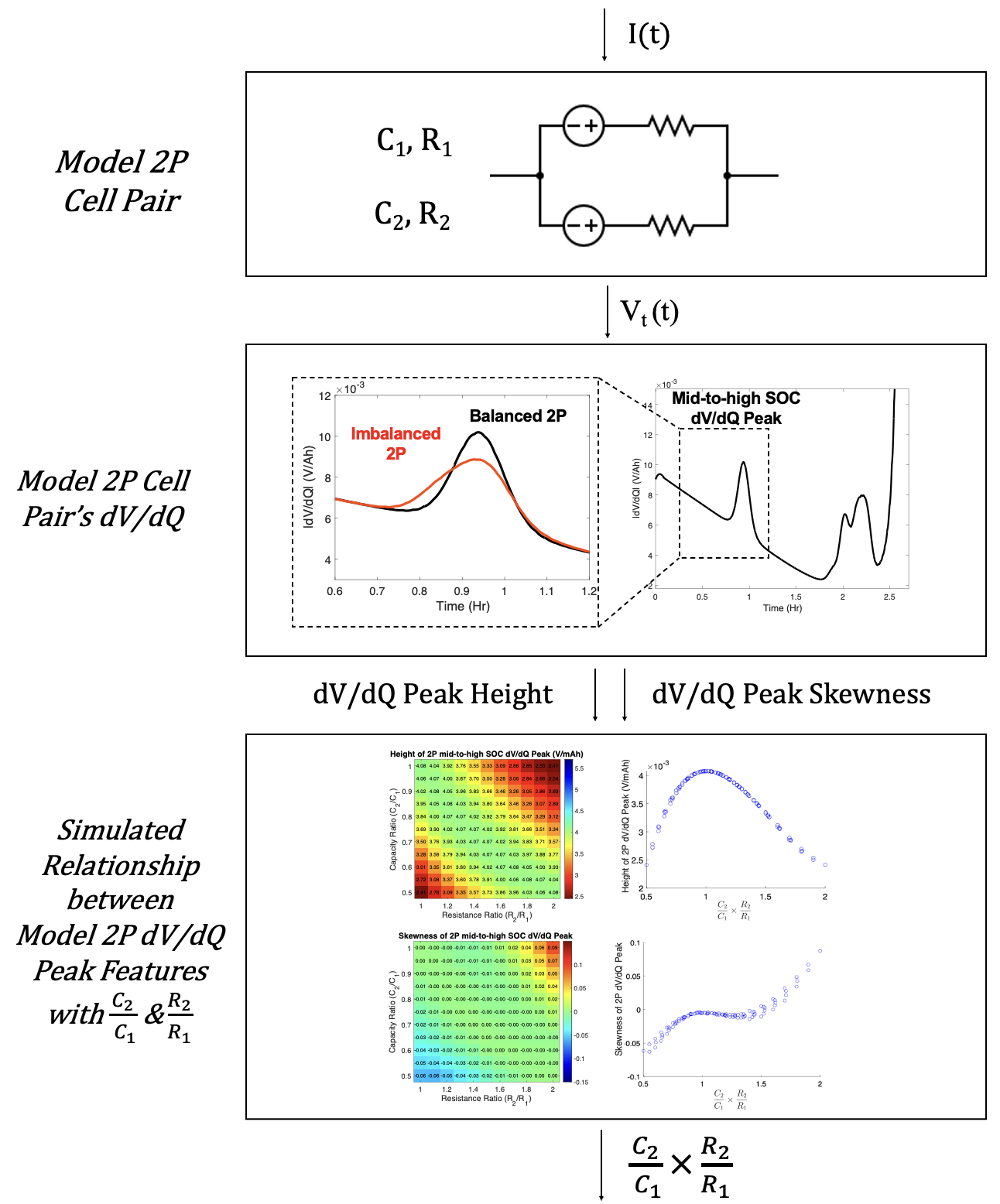}
    \caption{Flow diagram of our proposed method useful for identifying imbalances in capacity and resistance within a parallel-connected cell pair using only the pair's voltage, $V_t(t)$, and current, $I(t)$. The method leverages the relationship between features of the pair's dV/dQ peak and imbalances in capacity and resistance. 
    }
    \label{fig:Figure1_ACC_flow_diagram_16}
\end{figure}

This work introduces a novel method useful for identifying imbalances in capacity and resistance within a parallel-connected cell pair using only the pair's voltage and current measurements. Our method utilizes differential voltage analysis (DVA) when the pair is under constant current (CC) discharge and demonstrate that the features derived from the pair's differential voltage curve (dV/dQ), namely the mid-to-high SOC dV/dQ peak's height and skewness, are sensitive to imbalances in capacity and resistance. We establish the relationship between parallel-connected cell pair's dV/dQ peak shape features and imbalances in capacity and resistance by running various numerical simulations of a model parallel-connected cell pair undergoing a C/3 constant current (CC) discharge, where in each simulation the model parallel-connected cell pair has a unique combination of individual cell capacities and resistances. Ultimately, from this relationship, we demonstrate that dV/dQ peak shape features can identify the product of capacity imbalance and resistance imbalance but cannot uniquely identify the imbalances. (Fig. \ref{fig:Figure1_ACC_flow_diagram_16})

\section{Model Description}
Our model parallel-connected pair used in our simulations integrates equivalent circuit modeling with the intra-cycle dynamics of two parallel-connected cells. Subsection \ref{subsec:Dual-Tank Cell Model} outlines the model for the individual cells in a 2P cell pair, where the capacity and resistance of each individual cell are its parameters. Subsection \ref{subsec:Current Dynamics Description} describes the intra-cycle dynamics of two parallel-connected cells, presenting how differences in capacities and resistances among the cells affect each cell’s voltage and current.

\subsection{Model for Individual Cells in 2P Group}
\label{subsec:Dual-Tank Cell Model}
Each individual cell, $i$, is modeled as an OCV-R model. 
\begin{align}
    \label{eqn:cell Vt}
    V_{t,i}(t) &= OCV_i(z_i(t)) + I_i(t)\cdot R_i.
\end{align}
$V_{t,i}$ is the cell's output terminal voltage. $OCV_i$ is the cell's open circuit voltage with respect to $z_i(t)$,  the cell's SOC. $I_i(t)$ is the cell's input current, which is defined to be negative on discharge. $R_i$ is the cell's Ohmic resistance. The cell's SOC dynamics are described as follows:
\begin{equation}
    \label{eqn:zdot}
    \dot{z}_i(t) = \frac{1}{C_i} I_i(t).
\end{equation}
where $C_i$ is the cell's capacity. $R_i$ and $C_i$ are constant over the course of a single cycle.


$OCV_i$ is equal to the potential difference between the positive electrode $(U_p)$ and the negative electrode $(U_n)$: 
\begin{equation}
\label{eqn:OCV_OCP}
OCV_i(z_i(t)) = U_{p,i}(z_i(t)) - U_{n,i}(z_i(t))
\end{equation}

In this work, the model for individual cells is based on NMC/graphite chemistry. The NMC positive electrode potential \( U_{p,i}(z_i(t)) \) and graphite negative electrode potential \( U_{n,i}(z_i(t)) \) are shown in Fig. \ref{fig:OCP-OCV-function-v7}A and are written in Appendix B. Their corresponding differential voltages with respect to discharge coulomb counting, $Q_i$, are in  Fig. \ref{fig:OCP-OCV-function-v7}B. 

Phase transitions occur in graphite and cause sharp changes in \( U_{n,i}(z_i(t)) \), which more evidently result in peaks in \( \frac{dU_{n,i}}{dQ_{i}} \). These phase transitions consequently cause peaks in \( \frac{dV_{t,i}}{dQ_{i}} \) when \(I_i(t)\)  is low dynamic. The \( \frac{dV_{t,i}}{dQ_{i}} \) peaks, indicative of the phase transitions in graphite, provide insight into the individual cells' state of lithiation during discharge. In this work, the \( \frac{dV_{t,i}}{dQ_{i}} \) peaks from individual cells assist in explaining the changes in the parallel cell group's mid-to-high SOC \(dV/dQ\) peak shape caused by  imbalances in parallel-connected cells.

\begin{figure}[ht!]
    \centering
    \includegraphics[width=\linewidth]{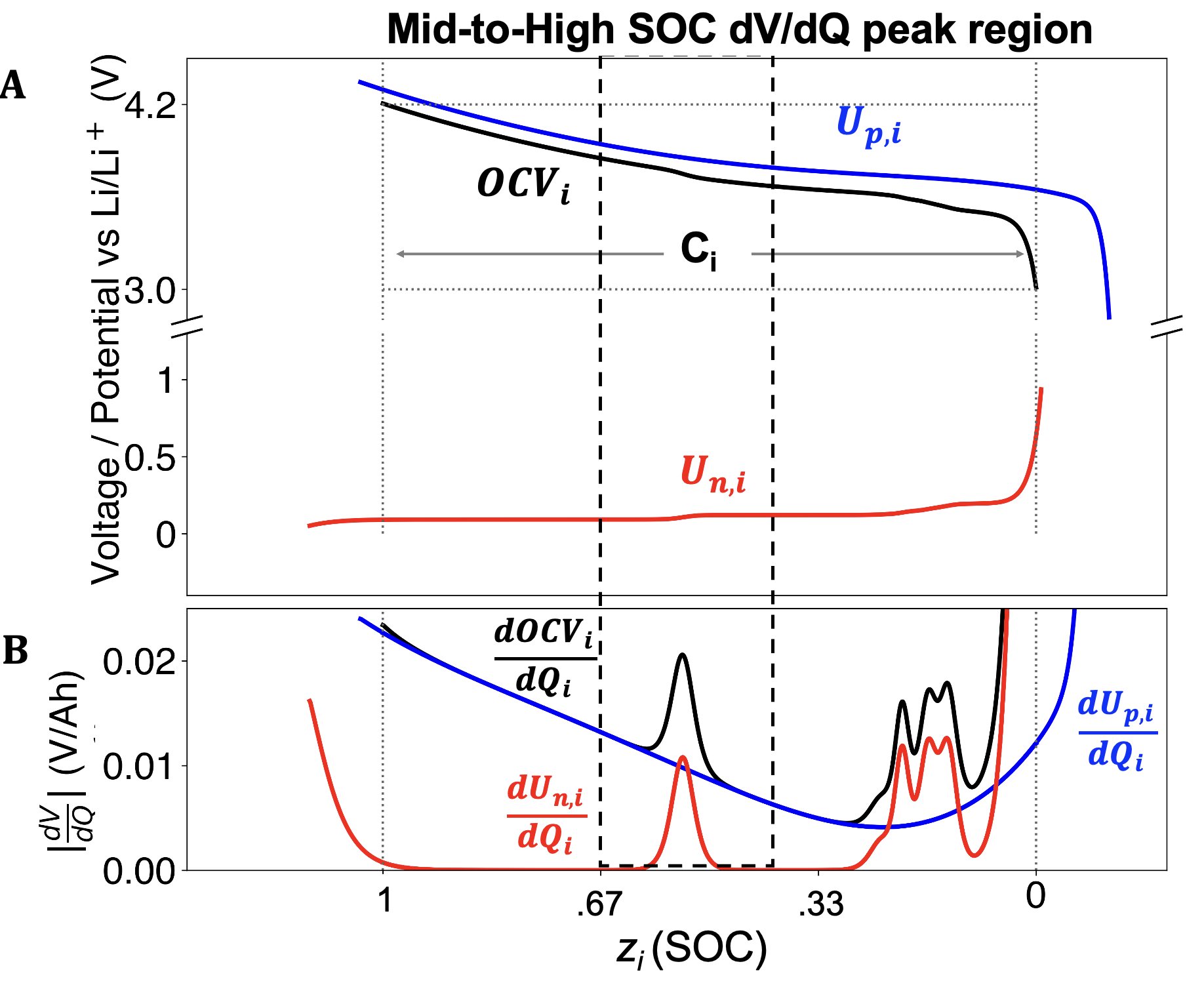}
    \caption{(Top) Full Cell Open Circuit Voltage ($OCV_i$) and Half-Cell Potentials ($U_{p_i}$, $U_{n_i}$) of NMC/graphite. (Bottom) Differential Voltage of the Full Cell and Half Cells with respect to Discharge Coulomb Counting,  $Q_i$. Figure adapted from \cite{Weng2023-lj}
    }
    \label{fig:OCP-OCV-function-v7}    
\end{figure}

\subsection{Current Split Dynamics in Parallel Imbalanced Groups}
\label{subsec:Current Dynamics Description}

This subsection outlines the intra-cycle dynamics of two parallel-connected cells implemented in the simulations. The dynamics are derived by Weng et al. \cite{Weng2022-pt}. Given that each cell within the 2P cell pair follows the OCV-R dynamics outlined in Eqs. \ref{eqn:cell Vt}- \ref{eqn:OCV_OCP}, the intra-cycle dynamics are as follows:
\begin{align}
    \label{eqn:vt}
    \begin{split}
    V_t(t) &= \frac{R_1\cdot OCV_2(z_2(t)) + R_2\cdot OCV_1(z_1(t))}{\rtot} \\
    &\quad + \frac{R_1R_2\itot(t)}{\rtot}
    \end{split}
    \\
    \label{eqn:kcl}
    \itot(t) &= I_1(t) + I_2(t) 
    \\
    \label{eqn:ib}
    I_1(t) &= \frac{-\Delta OCV(t) - R_2\itot(t)}{\rtot} 
    \\
    \label{eqn:ia}
    I_2(t) &= \frac{+\Delta OCV(t) - R_1\itot(t)}{\rtot}
\end{align}
where $V_{t}$ is the 2P cell pair's output terminal voltage, $I$ is the 2P cell pair's input current, $ \rtot \triangleq R_1 + R_2$, and $\Delta OCV(t) \triangleq OCV(z_2(t)) - OCV(z_1(t))$. Eqs. \ref{eqn:vt}- \ref{eqn:ia} show how imbalances in capacity and resistance within two parallel-connected cells affect the terminal voltage and current of the individual cells in a 2P cell pair.

\section{Simulations of 2P cell pair with capacity imbalance and resistance imbalance }

In this section, we present various numerical simulations of a 2P cell pair undergoing a C/3 CC discharge to show the relationship between features derived the pair’s differential voltage curve (dV/dQ), namely its mid-to-high SOC dV/dQ peak's height and skewness, and imbalances in capacity and resistance within parallel-connected cells. 

In Subsections \ref{subsec:Variability in capacity to shape} and \ref{subsec:Variability in resistance to shape}, we first examine the effects of capacity imbalance and resistance imbalance, respectively, on the dynamics of a 2P cell pair and its individual cells. We highlight that the effects of the imbalances are observable at the pair's dV/dQ peak, demonstrating the potential to identify imbalances within 2P cell pair through the pair's voltage and current measurements. We analyze and explain how and why the dV/dQ peak changes in response to these imbalances, highlighting that underlying current imbalance dynamics contribute to these changes. In Subsection \ref{subsec:Quantification of relationship on dV/dQ}, we present the sensitivity of the dV/dQ peak's height and skewness to capacity imbalance and resistance imbalance, independently. Finally, in Subsection \ref{subsec:Variability in capacity and resistance to shape}, we study the relationship between dV/dQ peak shape features with capacity imbalance and resistance imbalance jointly.


In this paper, we focus on the mid-to-high SOC parallel cell group dV/dQ peak. This peak typically occur in the voltage range of 3.7 - 3.9 V for NMC/graphite. To capture features of this peak, our initial step involves downselecting the voltage and current data corresponding to the 3.7 - 3.9 V range. The dV/dQ peak height can be estimated using MATLAB's \( \texttt{findpeaks} \) function applied to the downselected dV/dQ data. Subsequently, the dV/dQ peak skewness can be estimated following the algorithm described in Appendix A.

Table \ref{model_parameters} lists the model parameters of the individual cells for the simulations presented in this section. 



\begin{table}[h]
    \centering
    \caption{Model Parameters for Figures 3-6
    \newline $\alpha$ = ($C_2$/$C_1$),  $\beta$ = ($R_2$/$R_1$)}
    \label{model_parameters}
    \scriptsize
    \setlength{\tabcolsep}{3pt}
    \begin{tabularx}{\columnwidth}{ccccccccc}
        \toprule
        Figure & Group & Individual Cell & Capacity (Ah) & Resistance (m$\Omega$) \\
        \midrule
        3 & Bal. 2P & Bal   & 60 & 2 \\
          & Imb. 2P & Weak  & 40 & 2 \\
          &         & Strong & 80 & 2 \\
        \midrule
        4 & Bal. 2P & Bal   & 60 & 2 \\
          & Imb. 2P & Weak  & 60 & 3 \\
          &         & Strong & 60 & 1.5 \\
        \midrule
        5,6 & Imb. 2P & Weak & $60*\alpha/(\alpha+1)$ & $2*(\beta+1)$ \\
            &         & Strong & $60/(\alpha+1)$ & $2*(\beta+1)/\beta$ \\
        \bottomrule
    \end{tabularx}
\end{table}

For each simulation of an imbalanced 2P cell group, a strong cell (with higher capacity or lower resistance) is connected in parallel with a weak cell (with lower capacity or high resistance). Taking Cell 1 to be the strong cell and Cell 2 to be the weak cell, we thus impose
the constraint that $C_2$ $<$ $C_1$ and $R_2$ $>$ $R_1$.

In every simulation, the model 2P cell group has a total capacity of 120 Ah and a total resistance of 1 m$\Omega$. These values for capacity and resistance are realistic in real-world applications, particularly in energy storage systems.

\subsection{Effect of capacity imbalance on 2P cell pair's dynamics}

\label{subsec:Variability in capacity to shape}

\begin{figure}[ht!]
    \centering

    \includegraphics[width=\linewidth]{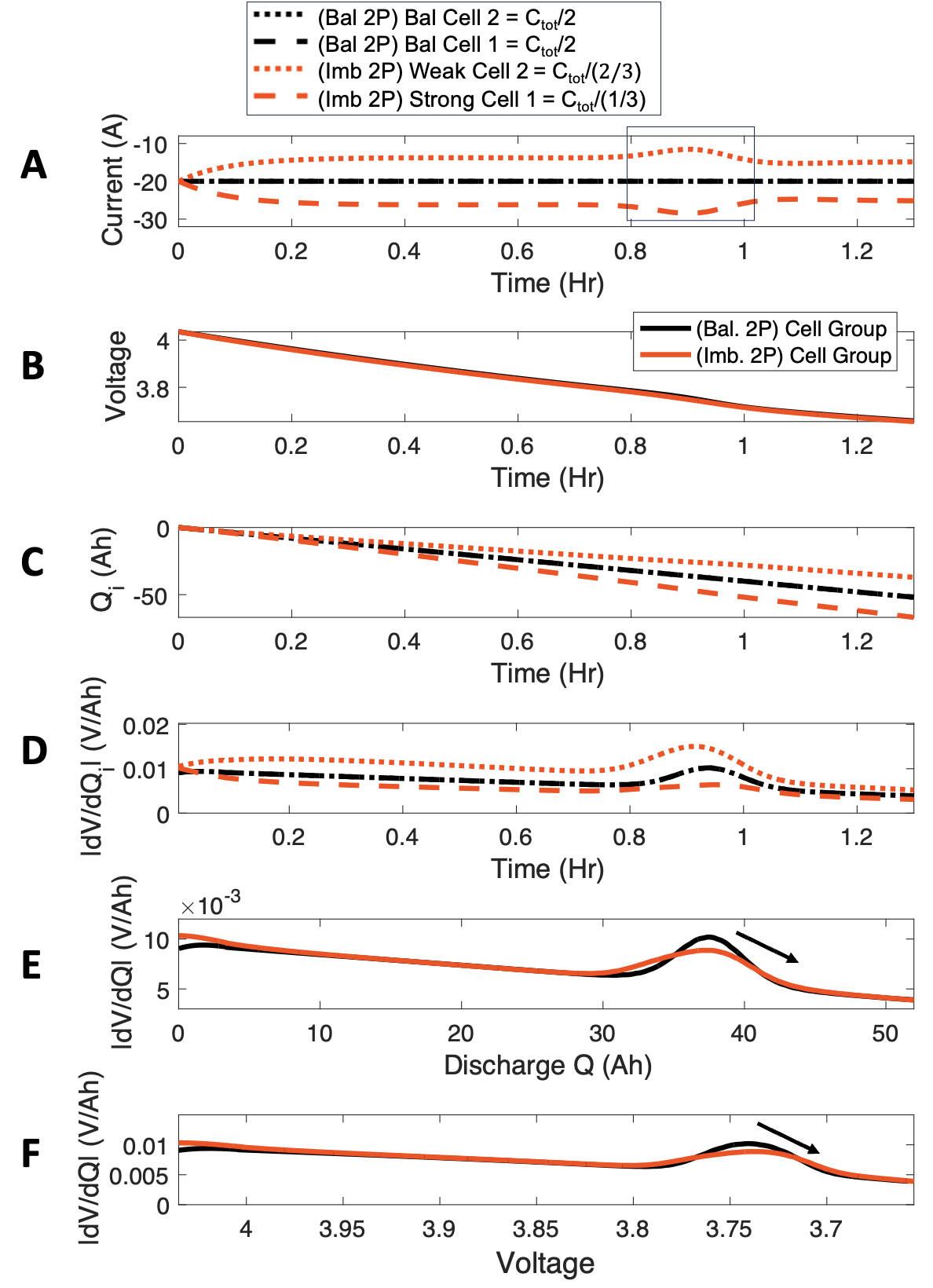}
    \caption{Simulation of a Balanced 2P Cell Group and of an Imbalanced 2P cell group with capacity imbalance ($C_2$/$C_1$ =.5) under C/3 constant current (CC) discharge. A) Individual Cell Current ($I_i$) vs Time, B) 2P Voltage  ($V_t$) vs Time, C) Individual Cell Discharge Coulomb Counting ($Q_i$) vs Time, D) Individual Cell dV/dQ ($\frac{dV}{dQ}_i$) vs Time, E) 2P dV/dQ ($\frac{dV}{dQ}$) vs 2P Discharge Coulomb Counting (Q), F) 2P dV/dQ ($\frac{dV}{dQ}$) vs 2P Voltage  ($V_t$)
    }
    
    \label{fig:C-variation-9}    
\end{figure}

To illustrate the effect of capacity imbalance on the dynamics of a 2P cell pair, we compare a simulation of 2P cell pair with balanced cells—those having identical capacity and resistance—to a simulation of 2P cell pair whose individual cells vary in capacity by a ratio of 0.5 ($C_2$/$C_1$ = 0.5) but have the same resistance (Fig. \ref{fig:C-variation-9}). 

Figure  \ref{fig:C-variation-9} presents the dynamics of 2P cell pairs and their individual cells for these simulations when the pairs are under C/3 constant current (CC) discharge. The comparison of the solid lines, representing the 2P cell pairs' dynamics, reveals that the effect of capacity imbalance within parallel-connected cells is particularly observable at the peak of the pair's dV/dQ curves, as illustrated in Fig. \ref{fig:C-variation-9}E \& F. This indicates that the pair's dV/dQ peak provides an opportunity estimate capacity imbalance using only the pair's voltage and current measurements. 

The pair's dV/dQ peak can be characterized by its height and skewness, and these simulations show that capacity imbalance causes the dV/dQ peak height to decrease and skew toward lower voltages (Fig. \ref{fig:C-variation-9}F). The effect of capacity imbalance on the pair's dV/dQ peak stems from individual cell current imbalances in the mid-to-high SOC graphite phase transition region. In this region, sharp changes in the slope of the individual cells' OCV with respect to their SOC drive non-linear oscillations in current. As the cells discharge through this SOC region, the strong cell with higher capacity experiences an increase in current, while the weak cell with lower capacity experiences a decrease in current (Fig. \ref{fig:C-variation-9}A). These oscillations affect the parallel cell group $V_t$ (Fig. \ref{fig:C-variation-9}B) and individual cell coulomb counting (Fig. \ref{fig:C-variation-9}C), but most evidently affect the individual cell \(\frac{dV}{dQ_{i}}\) peaks (Fig. \ref{fig:C-variation-9}D). The current oscillations cause the individual cell \(\frac{dV}{dQ_{i}}\) peaks to occur at different times during discharge. The timing of these peaks ultimately causes the pair's dV/dQ peak height to reduce and skew towards lower voltages.



\subsection{Effect of resistance imbalance on 2P cell pair's dynamics}
\label{subsec:Variability in resistance to shape}
\begin{figure}[ht!]
    \centering
    
    \includegraphics[width=\linewidth]{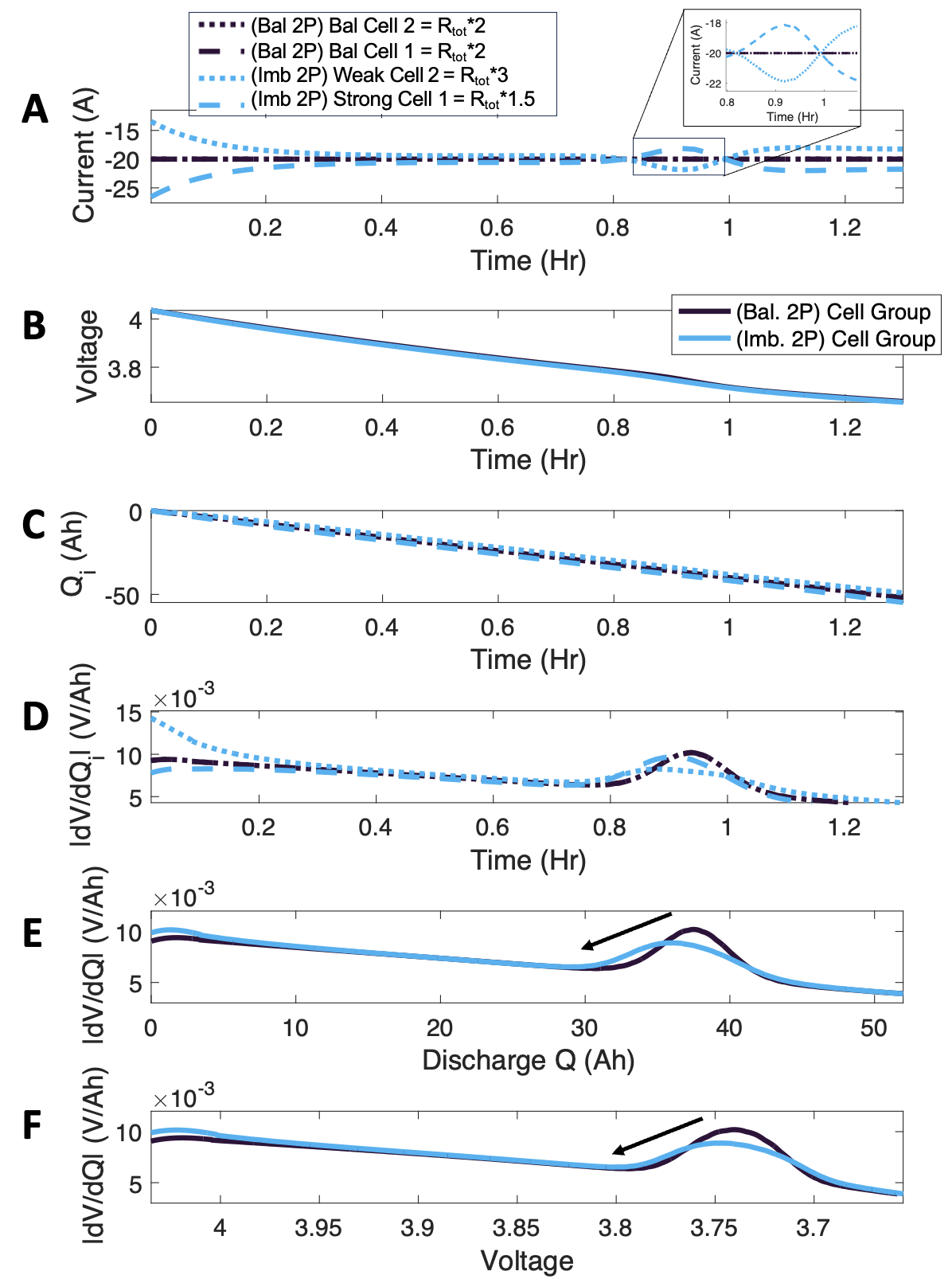}
    \caption{Simulation of a Balanced 2P Cell Group and of an Imbalanced 2P Cell Group with resistance imbalance ($R_2$/$R_1$ =2) under C/3 CC discharge. A) Individual Cell Current ($I_i$) vs Time, B) 2P Voltage  ($V_t$) vs Time, C) Individual Cell Discharge Coulomb Counting ($Q_i$) vs Time, D) Individual Cell dV/dQ ($\frac{dV}{dQ}_i$) vs Time, E) 2P dV/dQ ($\frac{dV}{dQ}$) vs 2P Discharge Coulomb Counting (Q), F) 2P dV/dQ ($\frac{dV}{dQ}$)  vs 2P Voltage  ($V_t$)}


    \label{fig:R-variation-9}
   
\end{figure}

We next demonstrate the effect of resistance imbalance on the dynamics of a 2P cell pair by comparing a simulation of a 2P cell group with balanced cells to simulation of a 2P group with imbalanced cells that vary in resistance by a ratio of 2 ($R_2$/$R_1$ = 2) but have the same 
capacity. Figure  \ref{fig:C-variation-9} presents the dynamics of 2P cell pairs and their individual cells for these simulations when the pairs are under C/3 CC discharge. Similarly to subsection \ref{subsec:Variability in capacity to shape}, the effect of resistance imbalance is particularly observable at the peak of the pair's dV/dQ curves, as illustrated in Fig. \ref{fig:R-variation-9} E \& F.

Resistance imbalance causes the pair's dV/dQ peak to decrease, similar to capacity imbalance; however, unlike capacity imbalance, it causes the peak to skew toward higher voltages (Fig. \ref{fig:R-variation-9}F) The changes in the dV/dQ peak are again driven by current oscillations near in mid-to-high SOC graphite phase transition region. However, the current oscillations differ: the weak cell with greater resistance experiences a decrease in current, while the strong cell with less resistance has an increase (Fig. \ref{fig:R-variation-9}A). These differing current oscillations consequently cause individual cell dV/dQ peaks to behave differently. As shown in Fig. \ref{fig:R-variation-9}D, a time gap still exists between the individual cell dV/dQ peaks, contributing to the group's dV/dQ peak broadening and a reduction in height. However, both individual dV/dQ peaks occur earlier, which means they arise at a higher 2P group voltage, ultimately causing the group's dV/dQ peak to skew toward higher voltages (Fig. \ref{fig:R-variation-9}F).


\subsection{Sensitivity of 2P dV/dQ peak shape features to capacity imbalance and resistance imbalance, independently}
\label{subsec:Quantification of relationship on dV/dQ}
\begin{figure}[ht!]
    \centering
    \includegraphics[width=\linewidth]{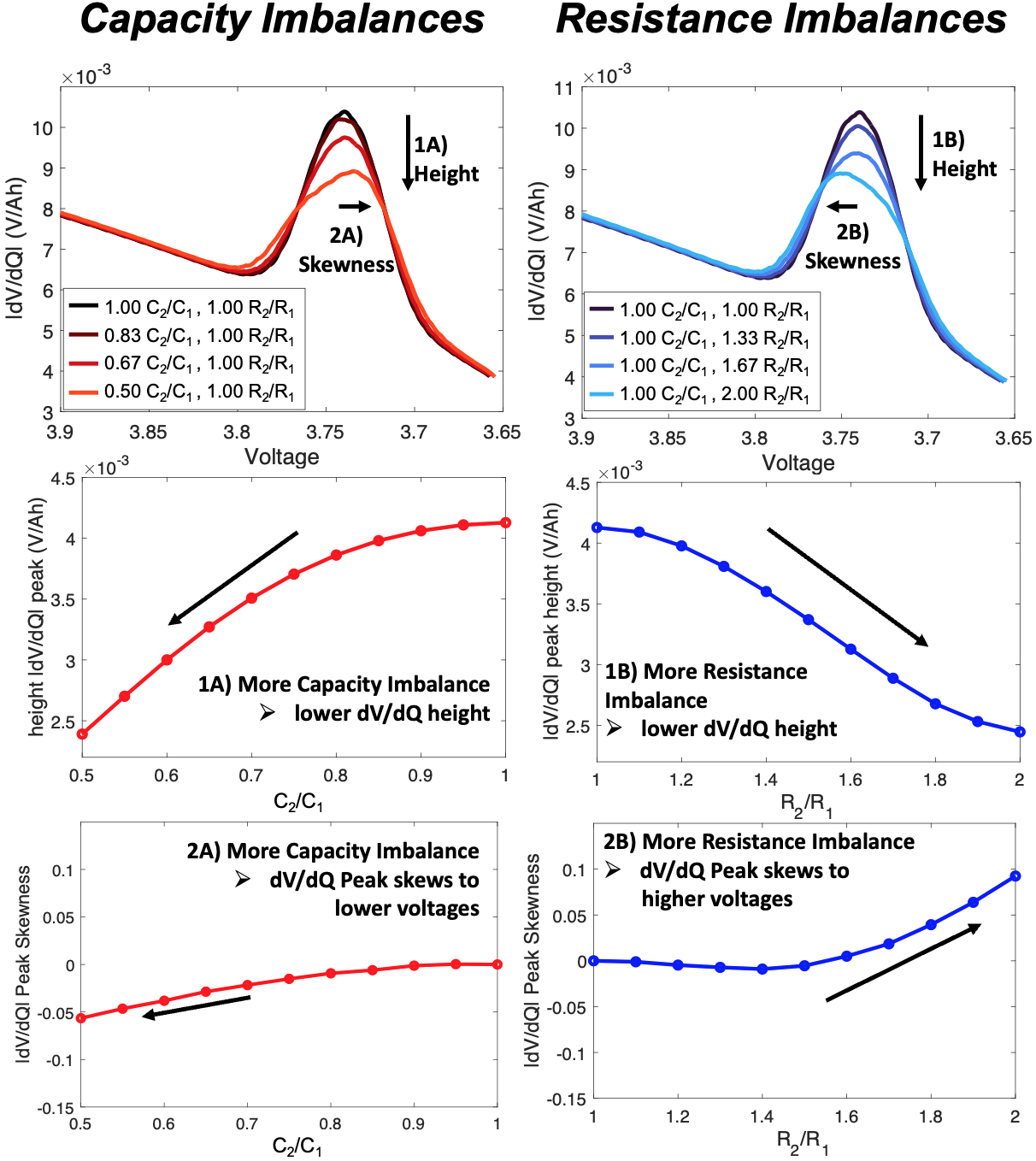}
    \caption{
    (Top) Comparison of 2P mid-to-high SOC dV/dQ peak vs 2P Voltage for varying capacity ratios ($C_2$/$C_1$) on the left and resistance ratios ($R_2$/$R_1$) on the right.
    (Middle) Height of 2P mid-to-high SOC dV/dQ vs $C_2$/$C_1$ on the left and $R_2$/$R_1$  on the right. 
    (Bottom) Skewness of 2P mid-to-high SOC dV/dQ vs $C_2$/$C_1$ on the left and vs $R_2$/$R_1$ on the right.  
    }
    \label{fig:figure_3_v6}
\end{figure}

Fig. \ref{fig:figure_3_v6} presents the sensitivity of 2P dV/dQ peak shape features to capacity imbalances and resistance imbalances, independently. It presents the 2P dV/dQ peak shape features as the capacity ratio ($C_2$/$C_1$) varies from 1 to 0.5 with the resistance ratio held constant, and as the resistance ratio ($R_2$/$R_1$) changes from 1 to 2 with the capacity ratio held constant. It indicates that as capacity imbalance increases, the 2P dV/dQ peak proportionally reduces in height and skews toward higher voltages. Meanwhile, as resistance imbalance increases, the 2P dV/dQ peak proportionally reduces in height but skews toward higher voltages.

\subsection{Sensitivity of 2P dV/dQ peak shape features to the joint effect of capacity imbalance and resistance imbalance}

\label{subsec:Variability in capacity and resistance to shape}

From \ref{subsec:Variability in capacity to shape} to \ref{subsec:Quantification of relationship on dV/dQ}, we have shown how capacity imbalance and resistance imbalance each independently affect 2P cell pair's dynamics and the sensitivity of specific features derived from the 2P dV/dQ curves, namely the mid-to-high SOC dV/dQ peak height and skewness, to these imbalances. In reality, however, parallel cell groups have capacity imbalance and resistance imbalance concurrently. In this subsection, we study the combined effects of capacity imbalance and resistance imbalance on mid-to-high SOC dV/dQ peak height and skewness.

\begin{figure}[ht!]
    \centering
    \includegraphics[width=\linewidth]{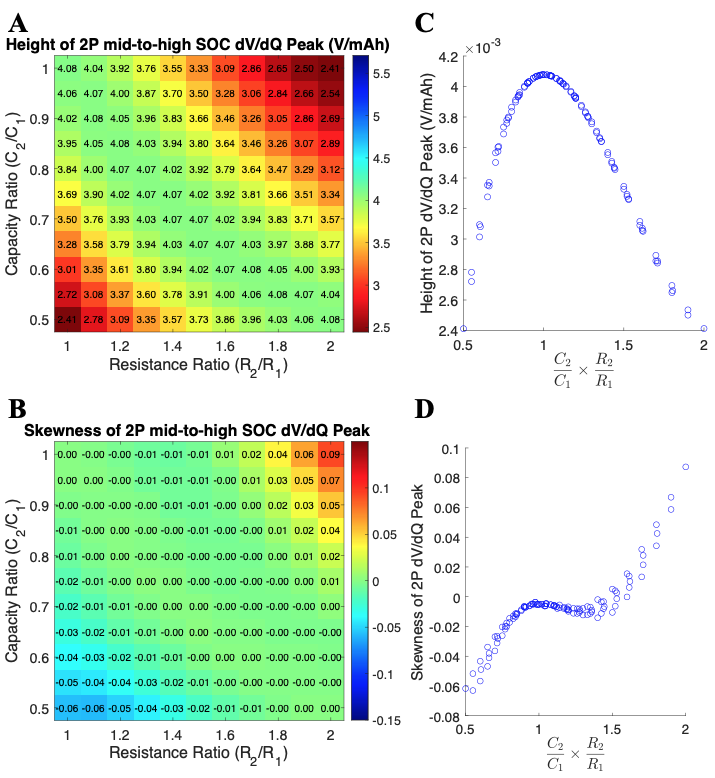}
    \caption{Heatmaps of the sensitivity of the height (A) and skewness (B) of the 2P mid-to-high SOC dV/dQ Peak to the joint effect of capacity imbalance and resistance imbalance. Scatter plots of the sensitivity of height (C) and skewness (D) of the 2P mid-to-high SOC dV/dQ Peak to the product of capacity ratio ($C_2$/$C_1$) and resistance ratio ($R_2$/$R_1$). $C_2$/$C_1$ varies from .5 to 1 and $R_2$/$R_1$ varies from 1 to 2.
    }
    \label{fig:heatmap_variability_in_cap_and_r_v9}
\end{figure}

Fig. \ref{fig:heatmap_variability_in_cap_and_r_v9} A \& B illustrates the combined effects of capacity and resistance imbalance on 2P mid-to-high SOC dV/dQ peak height and skewness, respectively. It aligns with observations in \ref{subsec:Variability in capacity to shape}, \ref{subsec:Variability in resistance to shape}, and \ref{subsec:Quantification of relationship on dV/dQ}. However, it additionally highlights that combinations of capacity ratio and resistance ratio cannot be uniquely identified based on mid-to-high SOC dV/dQ peak shape features. This is particularly evident when the product of the capacity ratio and resistance ratio is 1. In such cases, the non-linear oscillations in current, which primarily drive the changes in the mid-to-high SOC dV/dQ peak shape, are nullified \cite{Weng2022-pt} and there are no changes in the mid-to-high SOC dV/dQ peak shape. 

The product of capacity ratio and resistance ratio can be uniquely identified from the mid-to-high SOC dV/dQ peak shape features. Fig. \ref{fig:heatmap_variability_in_cap_and_r_v9} C \& D shows the sensitivity of dV/dQ peak shape features to the product of capacity ratio and resistance ratio. These figures illustrate that dV/dQ peak shape features can uniquely identify the product of capacity ratio and resistance ratio, particularly when the product deviates significantly from 1. As indicated in Figure \ref{fig:heatmap_variability_in_cap_and_r_v9} C, for a given value of dV/dQ Peak Height, there are two possible values for the product of capacity ratio and resistance ratio. The dV/dQ peak skewness differentiates the two potential values; a higher skewness value identifies the higher product of capacity ratio and resistance ratio, while a lower skewness value identifies the lower product of capacity ratio and resistance ratio. It is important to note, however, that when the product of capacity ratio and resistance ratio is close to 1, the dV/dQ skewness is not highly sensitive to the changes in the product of capacity ratio and resistance ratio, which prevents the unique identification of the product of the capacity ratio and resistance ratio.

\section{CONCLUSIONS AND FUTURE WORK}


This paper presents how imbalances in capacity and resistance within a pair of parallel-connected cells can be identified using only voltage and current measurements from the pair by leveraging features of the pair's dV/dQ, namely the height and skewness of the mid-to-high SOC dV/dQ peak. We establish the relationship between parallel-connected cell pair's dV/dQ peak shape features and imbalances in capacity and resistance by running various numerical simulations of a model parallel-connected cell pair undergoing a C/3 constant current discharge, where in each simulation the model parallel-connected cell pair has a unique combination of individual cell capacities and resistances. We first present the effects of capacity imbalance and resistance imbalance, respectively, on the 2P cell pair's dynamics to demonstrate how the effects of imbalances are particularly observable at the pair's dV/dQ peak. We analyze and explain how and why the dV/dQ peak changes in response to these imbalances, highlighting that underlying current imbalance dynamics contribute to these changes. We then quantify the sensitivity of the mid-to-high SOC dV/dQ peak features, its height and skewness, to capacity and resistance imbalances independently.  Results show that as capacity imbalance increases, the 2P dV/dQ peak proportionally reduces in height and skews toward higher voltages.  Similarly, as resistance imbalance increases, the 2P dV/dQ peak also proportionally reduces in height but skews toward higher voltages. We finally studied the relationship between dV/dQ peak shape features and the joint effects of capacity and resistance imbalance. We demonstrate that dV/dQ peak shape features can identify the product of capacity imbalance and resistance imbalance but cannot uniquely identify the imbalances.

In future work, we will investigate incorporating information about the relationship between capacity and resistance for the two cells based on their degradation mechanisms. Adding this information has the potential to identify the imbalances in capacity and resistance within parallel-connected cells and detect risky degradation in parallel-connected cells.

Additionally, we will study the sensitivity of dV/dQ shape features to noise typical in field data to understand how uncertainties in dV/dQ shape features can affect estimations of imbalances. We will investigate methods to reduce uncertainty in dV/dQ shape features to enable the practical application of our methods to accurately and precisely estimate capacity and resistance imbalances within parallel-connected pairs in real-world battery modules. 

Lastly, we will investigate whether our methods leveraging the dV/dQ peak shape features can be used to identify imbalances in capacity and resistance within parallel cell configurations larger than two cells.

\addtolength{\textheight}{-12cm}   




\subsection{dV/dQ Peak Skewness Calculation}
\label{subsec:dV/dQ Peak Skewness Calculation}

\begin{algorithm}
\label{skewness Procedure}
\small
\caption{\textbf{dV/dQ Peak Skewness Calculation }}
\begin{algorithmic}

\State \textbf{Data:}  Discharge Amp-hours \( Q \) and \( V_{t} \) that includes mid-to-high SOC \( dV/dQ \) 
\State \textbf{Procedure:} 
\State 1. \textbf{Downselect} 2P Cell Group Voltage, \(V_t\), to \(3.7-3.9 \, \text{V}\).
\State 2. \textbf{Estimate} 2P positive electrode potential corresponding mid-to-high SOC \( dV/dQ \), \( P(Q) = \hat{a} + [\hat{b} \cdot Q + \hat{c} \cdot Q^2] \), by solving for the following optimization problem for \( \hat{\theta} = \{\hat{a},\hat{b},\hat{c},\hat{d},\hat{e},\hat{f}\} \)
\[
\begin{aligned}
    \hat{\theta} &= \underset{\theta}{\arg\min} \left\| a + [b \cdot Q + c \cdot Q^2] - \left[ d \cdot \tanh\left(\frac{Q-e}{f}\right) \right] - V_t \right\|^2
\end{aligned}
\]
\State 3. \textbf{Estimate} 2P negative electrode potential corresponding mid-to-high SOC \( dV/dQ \), \( N(Q) \), by computing the difference \( P(Q) \) and \( V_{t} \)
\[
N(Q) = P(Q) - V_{t}
\]
\State 4. \textbf{Differentiate}: Apply Savitzky-Golay filter to \( N(Q) \) and differentiate filtered \( N(Q) \) to estimate \( \frac{dN}{dQ}(Q) \). 
\State 5. \textbf{Normalize}
\[
\frac{dN}{dQ}_{norm}(Q) = \frac{\frac{dN}{dQ}(Q)}{\sum_{i=1}^{n} \frac{dN}{dQ}(Q_i) \, dQ}
\]
\State 6. \textbf{Filter} regions where \( \frac{dN}{dQ}_{norm}(Q) < 0.005 \) to excludes regions not associated with mid-to-high SOC \( dV/dQ \) peak 
\State 7. \textbf{Calculate skewness} of the filtered \( \frac{dN}{dQ}_{norm} (Q)\) using Fisher's moment coefficient of skewness:  
\[
\text{Skewness} = \frac{\mu_3}{\sigma^3} = \frac{\sum_{i=1}^{n} (Q_i - \mu)^3 \frac{dN}{dQ}_{norm}(Q_i) }{\sigma^3}
\]
\\
where $\mu = \sum_{i=1}^{n}  Q_i \cdot  \frac{dN}{dQ}_{norm}(Q_i) $, 
\\$\sigma = \sqrt{\sum_{i=1}^{n}  (Q_i - \mu)^2  \frac{dN}{dQ}_{norm}(Q_i) }$, 
\\

\end{algorithmic}
\end{algorithm}
\subsection{Half-cell Potential Functions}
\label{subsec:Half-cell Potential Functions}
{\small 

\begin{equation}
\begin{split}
U_{p,i}(z) = & 3.6674 - 0.0225z + 0.5619z^2 + 0.6329z^3 \\
        & - 0.1957z^4 + 0.1016z^5 \\
        & - 0.5623 \exp(95.102(1 - z) - 97.036).
\end{split}
\end{equation}

\


\begin{equation}
\begin{split}
U_{n,i}(z) = & 0.063 + 0.8 e^{-75(0.83z + 0.007)} - 0.012 \tanh \left( \frac{z - 0.15}{0.019} \right) \\
& - 0.012 \tanh \left( \frac{z - 0.19}{0.019} \right) - 0.004 \tanh \left( \frac{z - 0.27}{0.024} \right) \\
& - 0.009 \tanh \left( \frac{z - 0.23}{0.016} \right) - 0.0145 \tanh \left( \frac{z - 0.59}{0.024} \right) \\
& - 0.080 \tanh \left( \frac{z - 1.24}{0.066} \right)
\end{split}
\end{equation}

\section*{ACKNOWLEDGMENT}

This work was supported by LG Energy Solutions. 









\bibliographystyle{ieeetr}
\bibliography{main}




\end{document}